\documentclass[12pt,letterpaper]{article}
\usepackage[left=1in,top=1in,right=1in,nohead]{geometry}
\usepackage{enumerate}
\usepackage{graphicx}
\usepackage{amsmath}
\usepackage{amsthm}
\usepackage{mathrsfs}
\usepackage{amssymb}
\usepackage{multirow}
\usepackage{color}
\usepackage{bm}
\usepackage{comment}
\usepackage{subfig}
\usepackage{paralist}
\usepackage[shortlabels]{enumitem}
\usepackage[pdf]{pstricks}
\usepackage{pst-coil,pstricks-add,pst-plot}
\usepackage{jheppub}
\usepackage{customcommands}

\title{More of the Bulk from Extremal Area Variations}

\author[1,2]{Ning Bao,}
\author[3]{ChunJun Cao,}
\author[4]{Sebastian Fischetti,}
\author[5,6]{Jason Pollack,}
\author[4,7]{and Yibo Zhong}

\affiliation[1]{Computational Science Initiative, Brookhaven National Laboratory, Upton, New York, 11973}
\affiliation[2]{Center for Theoretical Physics and Department of Physics, 
     University of California, Berkeley, CA 94720}
\affiliation[3]{Joint Center for Quantum Information and Computer Science, University of Maryland, College Park, MD, 20742, USA}
\affiliation[4]{Department of Physics, McGill University, Montr\'eal, QC, H3A 2T8, Canada}
\affiliation[5]{Department of Physics and Astronomy, University of British Columbia, Vancouver, BC V6T 1Z1, Canada}
\affiliation[6]{Quantum Information Center and Department of Computer Science, University of Texas at Austin, TX 78712}
\affiliation[7]{The Niels Bohr Institute, University of Copenhagen, Blegdamsvej 17, DK-2100 Copenhagen Ø, Denmark}

\emailAdd{ningbao75@gmail.com}
\emailAdd{ccj991@gmail.com}
\emailAdd{fischetti@physics.mcgill.ca}
\emailAdd{jpollack@cs.utexas.edu}
\emailAdd{nxb643@alumni.ku.dk}

\abstract{
It was shown recently in~\cite{Bao_2019}, building on work of Alexakis, Balehowksy, and Nachman~\cite{AleBal17}, that the geometry of (some portion of) a manifold with boundary is uniquely fixed by the areas of a foliation of two-dimensional disk-shaped surfaces anchored to the boundary.  In the context of AdS/CFT, this implies that (a portion of) a four-dimensional bulk geometry can be fixed uniquely from the entanglement entropies of disk-shaped boundary regions, subject to several constraints.  In this Note, we loosen some of these constraints, in particular allowing for the bulk foliation of extremal surfaces to be local and removing the constraint of disk topology; these generalizations ensure uniqueness of more of the deep bulk geometry by allowing for e.g.~surfaces anchored on disconnected asymptotic boundaries, or HRT surfaces past a phase transition. We also explore in more depth the generality of the local foliation requirement, showing that even in a highly dynamical geometry like AdS-Vaidya it is satisfied.
}

\begin{document}

\maketitle

\section{Introduction}
\label{sec:intro}

Using AdS/CFT to shed light on quantum gravity requires us to understand how to rephrase questions in the bulk in terms of the boundary theory---that is, how does bulk physics emerge from the boundary?  At perhaps the most fundamental level, we need to understand how the bulk itself is encoded in the boundary.  Most progress in this direction has focused on mapping the effective operator algebra of a semiclassical bulk to the operator algebra of the boundary, ranging from the early HKLL approach to refined perspectives based on insights from quantum information and quantum error correction~\cite{HamKab05,HamKab06,Kab11,AlmDon14,DonHar16,FauLew17,Cotler_2019,Chen_2020}; see e.g.~\cite{Har18} for some review.

On the other hand, one might instead be interested in recovering the bulk spacetime metric itself from the boundary, rather than merely perturbative quantum fields on a fixed background (in the language of quantum error correction, we are interested in determining the code subspace to which a particular boundary state belongs, rather than the representations of operators on a particular choice of code subspace).  As first noted in~\cite{Van09,Van10}, the bulk geometrization of boundary entanglement provided by the Ryu-Takayanagi (RT) and Hubeny-Rangamani-Takayanagi (HRT) formulas~\cite{RyuTak06,HubRan07,LewMal13,DonLew16} suggests that the bulk geometry should be encoded within the entanglement structure of the boundary.  This observation has led to much progress in connecting the emergence of bulk gravitational physics to boundary entanglement; see e.g.~\cite{LasMcD13,FauGui13,BalCze13,BalCho13,MyeRao14,CzeDon14,SwiVan14,CzeLam14,
EngFis15,CzeLam16,Mos16,FauHae17,KabLif18,Swi09,Swi12,PasYos15,BaoPen18,MilVid18,
RoySar18,BaoPen19,HaeMin19,BaoCha19,CaoQi20,AgoCac20,JokPon20} for a sample of results including perturbative recovery of the bulk Einstein equations, reconstruction of sufficiently symmetric and low-dimensional bulk geometries, and the approximation of bulk geometries via tensor networks (see also~\cite{EngHor16,EngHor16b,EngFis17,EngFis17b,HerHor20} for approaches to reconstructing the bulk conformal metric from the structure of boundary correlators, rather than entanglement).

This question --- how to reconstruct a bulk metric from entanglement entropies of the boundary theory? --- was investigated by some of us in~\cite{Bao_2019}, relying on mathematical inversion techniques used and developed in~\cite{AleBal17}.  In short,~\cite{Bao_2019} addressed a closely-related purely geometric question: it showed that in any pseudo-Riemannian manifold~$M$ of dimension~$d \geq 4$ with boundary~$\partial M$, the areas of boundary-anchored spacelike two-dimensional extremal surfaces are sufficient to fix the geometry in any region of~$M$ reached by such surfaces, under appropriate assumptions\footnote{In addition, one of us studied potential connections between these surfaces, bulk reconstruction outside of the entanglement wedge, and Wilson loops in \cite{BCNU_2020}.}.  Our primary purpose in this Note is to soften some of these assumptions, obtaining a substantially more general result.  We then explore the generality of a key assumption in this refined result.

More precisely, the refined statement we shall obtain is as follows.  Let~$(M,g_{ab})$ be some geometry of dimension~$d \geq 4$ and arbitrary signature with boundary~$\partial M$, and assume that
\begin{enumerate}
    \item A portion~$\Rcal$ of~$M$ is foliated by a continuous family of spacelike, two-dimensional extremal surfaces~$\Sigma(\lambda^i)$ anchored to~$\partial M$ (where~$\lambda^i$ are~$d-2$ parameters labeling the surfaces); \label{assump:foliation}
    \item The~$\Sigma(\lambda^i)$ are weakly stable in the sense defined in~\cite{EngFis19}; and \label{assump:stable}
    \item The~$\Sigma(\lambda^i)$ are planar\footnote{In this paper, when we say a surface~$\Sigma$ is planar we mean that it can be covered with a single coordinate chart; equivalently, there exists a map~$\psi: \Sigma \to \mathbb{R}^2$.  Notably, this \textit{need not} mean that~$\Sigma$ has the topology of the plane.  When~$\Sigma$ is planar, we will choose the map~$\psi$ such that the image~$\psi(\Sigma) \subset \mathbb{R}^2$ is compact.}. \label{assump:planar}
\end{enumerate}
Then the geometry in~$\Rcal$ is uniquely fixed by the following data, hereafter referred to as the \textit{boundary data}:
\begin{enumerate}[(a)]
	\item The induced metric and extrinsic curvature of~$\partial M$; \label{bndry1}
	\item The boundary curves~$\partial \Sigma(\lambda^i)$ on which the~$\Sigma(\lambda^i)$ are anchored\footnote{Because the~$\Sigma(\lambda^i)$ are planar, each connected component of~$\partial \Sigma(\lambda^i)$ is homeomorphic to a circle.}; and \label{bndry2}
	\item The area functional~$A[\Sigma]$ which gives the area of the~$\Sigma(\lambda^i)$ and of arbitrary small extremal variations thereof. \label{bndry3}
\end{enumerate}

The case of natural interest in AdS/CFT takes~$(M,g_{ab})$ to be a four-dimensional asymptotically locally AdS (AlAdS) spacetime, since in that case HRT surfaces are two-dimensional.  Thus if we take the~$\Sigma(\lambda^i)$ to be HRT surfaces, their areas compute the entanglement entropies of the boundary regions enclosed by the curves~$\partial \Sigma(\lambda^i)$, and we conclude that the bulk metric in~$\Rcal$ is fixed by boundary entanglement entropies\footnote{Technically the argument, either of~\cite{Bao_2019} or here, requires~$\partial M$ to be a finite boundary, which is not the case for an AlAdS spacetime.  In the AlAdS context, we imagine regulating the AdS boundary with a UV cutoff in the conventional manner of holographic renormalization, so that the argument can then apply.}.  In principle, our results also allow for the discussion of bulk metric reconstruction in more than four bulk dimensions, though in that case the areas of two-dimensional extremal surfaces no longer correspond to boundary entanglement entropy data.  Consequently, we will often have the case of a four-dimensional AlAdS spacetime in mind when we need to invoke physical motivation, though we emphasize that our geometric result is independent not only of the dimension but even of the signature of the bulk geometry~$(M,g_{ab})$.

\begin{figure}[t]
\centering
\includegraphics[page=2,width=0.3\textwidth]{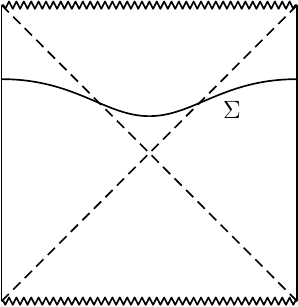}
\caption{An HRT surface~$\Sigma$ connecting two disconnected boundaries through a wormhole is not part of a family of surfaces that satisfies the assumptions of~\cite{Bao_2019}, and hence the argument of~\cite{Bao_2019} cannot be invoked to conclude that the metric in a neighborhood of~$\Sigma$ is uniquely fixed by boundary entanglement entropies.  The refined argument presented in this Note \textit{does} apply to~$\Sigma$, thereby allowing us to conclude that a portion of the black hole interior geometry is determined by boundary entanglement entropies.}
\label{fig:blackhole}
\end{figure}

We should highlight that our result is local: that is, given an HRT surface~$\Sigma$, the bulk metric in any sufficiently small neighborhood of~$\Sigma$ is fixed by boundary data, as long as that neighborhood can be foliated by (planar) HRT surfaces (constructed by e.g.~deforming~$\Sigma$ appropriately).  Importantly, this allows us to apply our result to cases like that shown in Figure~\ref{fig:blackhole}, in which HRT surfaces anchored to two disconnected boundaries go through an eternal black hole, to conclude that the metric in (part of) the black hole interior is uniquely fixed by boundary entanglement entropies; the assumptions required in~\cite{Bao_2019} exclude such surfaces.  Note, however, that we can say nothing about regions like the ``bulge'' of the recently discussed ``Python's lunch'' geometries~\cite{Brown_2020,bao2020warping}, nor the interiors of ``bag of gold'' geometries~\cite{Marolf_2009}, which are not reached by boundary-anchored HRT surfaces.  The reconstruction of such portions of the bulk would presumably require a different set of boundary data than those used here.

To describe our result, for convenience to the reader we begin with a review of the relevant aspects of~\cite{Bao_2019} in Section~\ref{sec:review}.  The main argument, using only Assumptions~\ref{assump:foliation},~\ref{assump:stable}, and~\ref{assump:planar} above, is described in Section~\ref{sec:generalization}.  The most nontrivial of these assumptions is that of the existence of the foliation~$\Sigma(\lambda^i)$, Assumption~\ref{assump:foliation}.  We explore its generality in Section~\ref{sec:foliation}, noting that it holds identically in static spacetimes, but also at least in certain highly-dynamical geometries like Vaidya-AdS.  We also make some comments on potential connections between the existence of such a foliation and properties of the Jacobi operators of the surfaces~$\Sigma(\lambda^i)$.  We conclude in Section~\ref{sec:disc} with some open questions.

\section{Review}
\label{sec:review}

In this Section we provide a broad-level review of the argument of~\cite{Bao_2019}, describing only those details necessary for following the arguments in Section~\ref{sec:generalization} below. In \cite{Bao_2019}, it was shown that if a pseudo-Riemannian manifold~$(M,g_{ab})$ of dimension~$d \geq 4$ with boundary~$\partial M$ satisfies the assumptions
\begin{enumerate}[label={\arabic*$'$}]
	\item[1.] A portion~$\Rcal$ of~$M$ is foliated by a continuous family of spacelike, two-dimensional extremal surfaces~$\Sigma(\lambda^i)$ anchored to~$\partial M$ (where~$\lambda^i$ are~$d-2$ parameters labeling the surfaces),
	\item[2.] The~$\Sigma(\lambda^i)$ are weakly stable,
	\setcounter{enumi}{2}
	\item The~$\Sigma(\lambda^i)$ are topological disks, and \label{assump:disk}
	\item The foliation~$\Sigma(\lambda^i)$ shrinks to a point on~$\partial M$ in an appropriate limit of the~$\lambda^i$, \label{assump:point}
\end{enumerate}
then the geometry of~$\mathcal{R}$ is uniquely fixed by the aforementioned boundary data~\ref{bndry1},~\ref{bndry2}, and~\ref{bndry3}.  In contrast with the assumptions listed in Section~\ref{sec:intro}, Assumption~\ref{assump:disk} is a notably stronger version of Assumption~\ref{assump:planar}, while Assumption~\ref{assump:point} is new.  Consequently, the argument of~\cite{Bao_2019} only fixes the metric in regions that can be reached by HRT surfaces that start very near~$\partial M$ and continuously move inwards; it could say nothing about what happens when HRT surfaces ``jump'', say due to a phase transition in the entanglement entropy.  Furthermore,~\cite{Bao_2019} required that the~$\Sigma(\lambda^i)$ all have disk topology, which excluded surfaces anchored to, say, two asymptotic boundaries, or more generally to any boundary region that does not also have disk topology. % As we see, the argument presented in this Note removes the first of these two requirements entirely, and lightens the second to merely that the~$\Sigma(\lambda^i)$ be planar.

\begin{figure}
\centering
\includegraphics[page=1,width=0.45\textwidth]{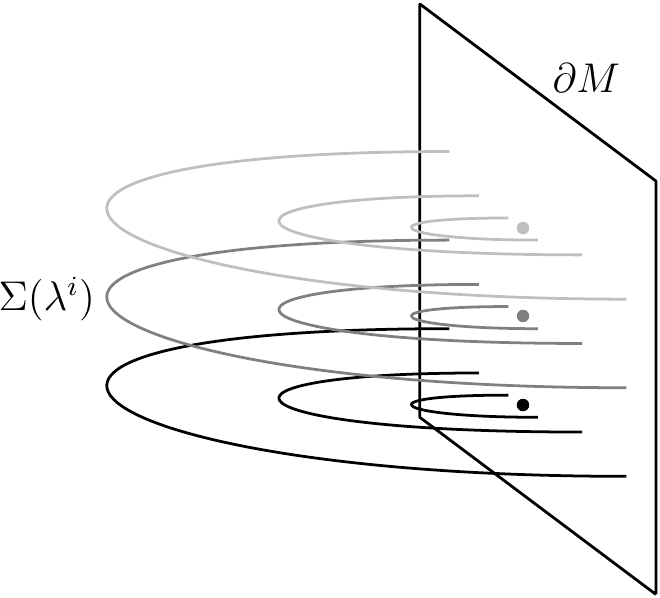}
\caption{The foliation~$\Sigma(\lambda^i)$ used in the argument of~\cite{Bao_2019}.  The~$\Sigma(\lambda^i)$ are a family of disk-shaped boundary-anchored extremal surfaces that foliate some portion of the bulk (here we suppress a spatial dimension, so the~$\Sigma(\lambda^i)$ appear as curves).  This family of surfaces degenerates to the marked points on the boundary.}
\label{fig:foliation}
\end{figure}

We shall take the indices of the parameters~$\lambda^i$ to take values~$i = 3, \ldots, d$.  A crucial fact is that linear deformations of extremal surfaces satisfy the Jacobi equation, which is a linear elliptic equation.  Specifically, consider some extremal surface~$\Sigma$, and extend this surface to a one-parameter family~$\Sigma(\lambda)$ of extremal surfaces, with~$\Sigma(0) = \Sigma$ (where for now~$\lambda$ is just a single parameter).  The deviation vector field~$\eta^a = (\partial_\lambda)^a$ can be taken normal to~$\Sigma$, and this normal component encodes the notion of an ``infinitesimal variation'' of~$\Sigma$ along the family~$\Sigma(\lambda)$.  Because the~$\Sigma(\lambda)$ are all extremal,~$\eta^a$ is constrained to obey the Jacobi equation
\begin{subequations}
\label{eq:Jacobi}
\begin{align}
0 = J_\Sigma \eta^a &\equiv -D^2 \eta^a - {Q^a}_b \eta^b, \\
Q_{ab} &\equiv K_{acd} {K_b}^{cd} + h^{cd} {P_a}^e {P_b}^f R_{ecfd}, \label{subeq:JacobiQ}
\end{align}
\end{subequations}
where~$D^2$ is the Laplacian on the normal bundle of~$\Sigma$,~$h_{ab}$ and~${K^a}_{bc}$ are the induced metric and extrinsic curvature of~$\Sigma$,~$P_{ab} \equiv g_{ab} - h_{ab}$ is the orthogonal projector to~$\Sigma$, and~$R_{abcd}$ is the Riemann curvature tensor of~$(M,g_{ab})$; see e.g.~\cite{EngFis19,Sim68,ColMin,LarFro93,Guv93,VisPar96,Car92,Car92b,Car93,BatCar95,BatCar00,Mos17,GhoMis17,LewPar18,Spe19} for a review and more explicit expressions.  It is sometimes convenient to decompose the Jacobi equation by introducing a basis~$\{(n^i)_a\}$,~$i = 3, \ldots, d$, of the normal bundle of~$\Sigma$, and then working with the components~$\eta^i \equiv (n^i)_a \eta^a$,~$J_\Sigma \eta^i \equiv (n^i)_a J_\Sigma \eta^a$.  In what follows we will take~$\{(n^i)_a\}$ to be the basis of coordinate one-forms~$\{(d\lambda^i)_a\}$.

With this in mind, the argument proceeds as follows.

\subsubsection*{Choice of Coordinate System}
\label{subsec:coords}

First, we make a convenient choice of coordinate system.  Because the~$\Sigma(\lambda^i)$ foliate~$\Rcal$, the~$(d-2)$ parameters~$\lambda^i$ are good coordinates in~$\Rcal$.  To complete the coordinate system, we exploit the fact that the~$\Sigma(\lambda^i)$ are two-dimensional and have disk topology to introduce conformally flat coordinates~$x^\alpha$ (with~$\alpha = 1,2$) on each~$\Sigma(\lambda^i)$; in these coordinates, the induced metric~$\sigma_{ab}$ on~$\Sigma(\lambda^i)$ takes the form
\be
\label{eq:isothermal}
\sigma_{\alpha\beta} dx^\alpha \, dx^\beta = e^{2\phi(x)} \left((dx^1)^2 + (dx^2)^2\right).
\ee
Now, such coordinates may always be introduced locally on any two-dimensional geometry, but in fact the boundary data allows the~$x^\alpha$ to be fixed globally in a way that preserves the boundary structure.  To explain this point in more detail, let us consider first introducing some arbitrary coordinate system~$\{y^\alpha\}$ on~$\Sigma$ (we suppress the foliation parameters~$\lambda^i$, since here we only need to consider a single surface at a time); these coordinates are defined by a map~$\psi: \Sigma \to \mathbb{R}^2$.  Two different metrics~$g_1$,~$g_2$ in~$\Rcal$ (and therefore different metrics~$\sigma_1$,~$\sigma_2$ on~$\Sigma$) will in general have different components in the~$\{y^\alpha\}$ coordinates, but the region~$\psi(\Sigma) \subset \mathbb{R}^2$ covered by the~$\{y^\alpha\}$ will be the same because the map~$\psi$ is independent of the metric.

\begin{figure}[t]
\centering
\includegraphics[page=3,width=0.9\textwidth]{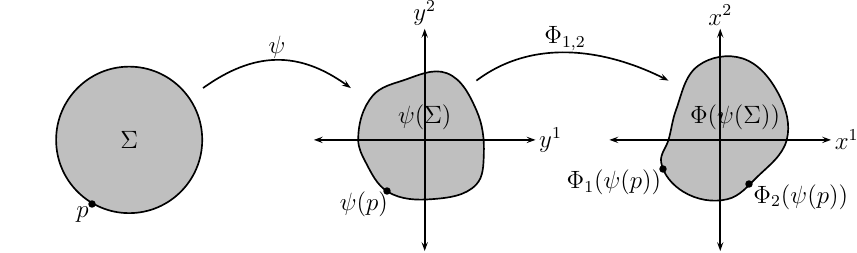}
\caption{Given two metrics~$\sigma_1$ and~$\sigma_2$ on~$\Sigma$, we may imagine introducing isothermal coordinates on~$\Sigma$ by first introducing a metric-independent coordinate system~$\{y^\alpha\}$ via a map~$\psi$, followed by the metric-dependent maps~$\Phi_1$ and~$\Phi_2$.  For general~$\sigma_1$ and~$\sigma_2$, these maps can be chosen to give the same image of~$\Sigma$ in the~$x^\alpha$-plane, but they need not agree pointwise; in particular, a point~$p \in \partial \Sigma$ need not correspond to the same~$\{x^\alpha\}$ coordinates via the two maps.}
\label{fig:isothermalmaps}
\end{figure}

To convert to the isothermal coordinates~$\{x^\alpha\}$, we must introduce another map~$\Phi$, which now \textit{does} depend on the metric; let us therefore denote by~$\Phi_1$ and~$\Phi_2$ two maps that put the metrics~$\sigma_1$ and~$\sigma_2$ on~$\Sigma$ in the form~\eqref{eq:isothermal}.  These maps are non-unique, and if we wish we may use the residual freedom to ensure that the images~$\Phi_1(\psi(\Sigma))$ and~$\Phi_2(\psi(\Sigma))$ coincide.  However, in general there is no guarantee that a boundary point~$p \in \partial \Sigma$ will have the same image under these two maps:~$\Phi_1(\psi(p)) \neq \Phi_2(\psi(p))$, as shown in Figure~\ref{fig:isothermalmaps}.  We would then be unable to compare the boundary data of~$g_1$ and~$g_2$ in the shared coordinate system~$\{x^\alpha\}$, and in particular we could not make use of the fact that the boundary data agrees.  Fortunately, it turns out that the agreement of the boundary data on~$\Sigma$ (and therefore also in the~$\{y^\alpha\}$ coordinates) can be used to ensure that the maps~$\Phi_1$ and~$\Phi_2$ \textit{do} coincide on the boundary:~$\Phi_1(\psi(p)) = \Phi_2(\psi(p))$ for~$p \in \partial \Sigma$; this is what we meant above when we said that it is possible to introduce isothermal coordinates that preserve the boundary structure.

To describe why this is the case, we first recall that the area functional~$A[\Sigma]$ of a particular boundary-anchored extremal surface~$\Sigma$ and of small extremal deformations thereof fixes some properties of the Jacobi operator~$J_\Sigma$ of~$\Sigma$.  Specifically, it fixes the \textit{Cauchy data}~$\Ccal_\Sigma$ of~$J_\Sigma$, which refers to the set of Dirichlet and Neumann boundary data consistent with solutions to the Jacobi equation:
\be
\Ccal_\Sigma = \left\{(\eta^i, N^b D_b \eta^i) | J_\Sigma \eta^i = 0\right\},
\ee
where~$N^a$ is the outward-pointing normal to~$\partial \Sigma$ in~$\Sigma$ (and as above,~$D^a$ is the covariant derivative on the normal bundle of~$\Sigma$, with~$N^b D_b \eta^i \equiv (d\lambda^i)_a N^b D_b \eta^a$)\footnote{Equivalently, one can interpret~$\Ccal_\Sigma$ as giving the Dirichlet-to-Neumann map~$\Psi: \eta^i \mapsto N^b D_b \eta^i$ from an arbitrary inhomogeneous Dirichlet boundary condition of the problem~$J_\Sigma \eta^i = 0$ to the boundary normal derivative~$N^b D_b \eta^i$ of the solution.}.  In short, knowing~$A[\Sigma]$ amounts to knowing~$\Ccal_\Sigma$.  With~$\Ccal_\Sigma$ in hand, we may then consider the ``exterior'' boundary-value problem
\begin{subequations}
\label{eqs:exterior}
\begin{align}
J_\Sigma \eta_\xi^i = 0 \mbox{ on } \Omega_y \equiv \mathbb{R}^2 \setminus \psi(\Sigma), \\
e^{-(y^1 + i y^2)\xi} \eta_\xi^i - \bar{\eta}^i \to 0 \mbox{ at large } |y|, \\
\left(\eta_\xi^i|_{\partial \psi(\Sigma)}, N^b D_b \eta^i_\xi|_{\partial \psi(\Sigma)}\right) \in \Ccal_\Sigma,
\end{align}
\end{subequations}
where~$\xi$ is an arbitrary nonzero complex number,~$\bar{\eta}^i$ are arbitrary coefficients, and~$J_\Sigma$ is extended to the entire~$(y^1, y^2)$ plane by taking~$Q_{ij}$ to vanish and~$D_a$ to be flat in the exterior region~$\Omega_y$ and taking~$\sigma_{\alpha\beta} = \delta_{\alpha\beta}$ outside of some region containing~$\psi(\Sigma)$, with~$\sigma_{\alpha\beta}$ fixed and known everywhere in~$\Omega_y$ (note that the letter~$i$ is playing double duty as both an index and as the imaginary unit; we assume it is clear from context which is which).  As argued in~\cite{Bao_2019} with logic reproduced in Section~\ref{subsec:topology} below, the problem~\eqref{eqs:exterior} has a unique solution for any~$\xi$ and~$\bar{\eta}^i$, and moreover the solution~$\eta_\xi^i$ exhibits the behavior
\be
\label{eq:etaxibound}
\left\|e^{-(x^1(y) + i x^2(y))\xi} \eta_\xi^i(y) - \bar{\eta}^i\right\|_{L^2(\Omega_y)} \to 0 \mbox{ as } |\xi| \to \infty \qquad \forall i = 3, \ldots, d,
\ee
where~$x^1(y)$,~$x^2(y)$ are the isothermal coordinates defined by the map~$\Phi$, fixed uniquely by requiring that~$(x^1(y), x^2(y)) \to (y^1, y^2)$ at large~$|y|$~\cite{Ahlfors}.  From this bound it follows that the maps~$\Phi_1$ and~$\Phi_2$ corresponding to different metrics~$\sigma_1$ and~$\sigma_2$ must in fact coincide everywhere in the exterior region~$\Omega_y$, including in particular on the boundary~$\partial \psi(\Sigma)$.  This can be seen intuitively from~\eqref{eq:etaxibound} by noting that the exterior problem~\eqref{eqs:exterior} is independent of the metric on~$\Sigma$ (as long as the Cauchy data~$\Ccal_\Sigma$ is fixed), and hence the solution~$\eta_\xi^i$ must be independent of the metric on~$\Sigma$ as well.  Since the left-hand side of~\eqref{eq:etaxibound} vanishes as~$|\xi|$ becomes large, the fact that~$\eta^i_\xi$ and~$\bar{\eta}^i$ are independent of the metric on~$\Sigma$ then implies that~$x^1(y) + i x^2(y)$ must be as well, at least for~$y \in \Omega_y$, and hence~$\Phi_1(y) = \Phi_2(y)$ for~$y \in \partial \psi(\Sigma)$.

The upshot is therefore that we may consistently choose to work in the isothermal coordinates without spoiling the structure of the boundary data; hence the coordinate system~$\{x^\alpha, \lambda^i\}$ is a good one to use in the region~$\Rcal$.

\subsubsection*{Fixing the Metric Functions $g^{ij}$ and $g^{i\alpha}$}

Once the coordinate system~$\{x^\alpha,\lambda^i\}$ is fixed, showing that the geometry in~$\Rcal$ is uniquely fixed by the available boundary data amounts to showing that the metric components in this gauge are fixed.  It is in fact easier to work with the inverse metric, so we need to fix~$g^{ij} \equiv g^{ab} (d\lambda^i)_a (d\lambda^j)_b$,~$g^{i\alpha} \equiv g^{ab}(d\lambda^i)_a (dx^\alpha)_b$, and the conformal factor~$\phi$.  The details of how the components~$g^{ij}$ and~$g^{i\alpha}$ are fixed are unimportant in what follows, so we will be especially brief.  The~$g^{ij}$ are fixed from the Cauchy data~$\Ccal_{\Sigma(\lambda^i)}$ via the uniqueness theorem of~\cite{AlbGui13}.  The~$g^{i\alpha}$ are fixed by considering a slight perturbation to the foliation~$\Sigma(\lambda^i)$ which ``tilts'' all the leaves of the foliation; this perturbation always exists by virtue of the fact that the~$\Sigma(\lambda^i)$ are weakly stable.  The ``tilted'' foliation corresponds to a slight perturbation of the coordinate system~$\{x^\alpha, \lambda^i\}$, and transforming the metric components between these two coordinate systems yields an invertible system of linear equations for the components~$g^{i\alpha}$ in terms of the boundary data.  Importantly, these metric components are obtained locally in the foliation: to obtain~$g^{ij}$ and~$g^{i\alpha}$ on a particular leaf~$\Sigma_*$ of the foliation, one needs access only to extremal surfaces that are small deformations of~$\Sigma_*$.

\subsubsection*{Fixing the Conformal Factor}

The final step consists of fixing the conformal factor~$\phi$ on each of the surfaces.  The extremality of the~$\Sigma(\lambda^i)$ can be expressed as~$(d-2)$ first-order linear PDEs obeyed by the conformal factor:
\be
\label{eq:phihyperbolic}
\sum_{\alpha=1}^2 (\partial_{\alpha} f_{\alpha i}+2f_{\alpha i}\partial_{\alpha}\phi)-2\partial_i\phi = 0 \qquad \forall i = 3, \ldots, d,
\ee
where~$f_{\alpha i} \equiv e^{-2\phi} g_{i\alpha}$ are functions only of~$g^{ij}$ and~$g^{i\alpha}$ (and not~$\phi$), and are therefore fixed by the boundary data.  Now we use the global property of the foliation, namely that the surfaces~$\Sigma(\lambda^i)$ degenerate to a point on the boundary~$\partial M$ for an appropriate limit of the parameters~$\lambda^i$.  Since~$\phi$ is known at~$\partial M$, we may therefore interpret~\eqref{eq:phihyperbolic} as~$d-2$ hyperbolic equations, with the parameters~$\lambda^i$ treated as time; by evolving~\eqref{eq:phihyperbolic} inwards from the boundary along an appropriate choice of~$\lambda^i$ (see Figure~\ref{subfig:oldevolution} below), we may then obtain~$\phi$ on any desired surface in the foliation.  Since the coefficients of~\eqref{eq:phihyperbolic} are fixed by boundary data (along with the boundary value of~$\phi$), we therefore find that the conformal factor is uniquely fixed everywhere in~$\Rcal$.  Thus all the metric components in~$\Rcal$ are fixed by boundary data, and the argument is complete.

\section{Generalizations}
\label{sec:generalization}

In this Section, we will relax two of the assumptions used in the argument reviewed above.  First, we will remove the restriction~\ref{assump:disk} that the surfaces~$\Sigma(\lambda^i)$ have disk topology, instead requiring $\Sigma(\lambda^i)$ merely be planar.  Second, we remove the restriction~\ref{assump:point} that the family~$\Sigma(\lambda^i)$ degenerate to a point on~$\partial M$ for an appropriate limit of the~$\lambda^i$.

\subsection{General Topology}
\label{subsec:topology}

In the original argument of~\cite{Bao_2019}, the assumption that the~$\Sigma(\lambda^i)$ have disk topology was used once: in deducing the existence of the isothermal coordinates~$\{x^\alpha\}$.  In fact, it is very straightforward to generalize the argument away from disk topology; the generalization proceeds essentially identically to that reviewed in Section~\ref{subsec:coords}.  For now we will again suppress dependence on the foliation parameters~$\lambda^i$ to write~$\Sigma$ rather than~$\Sigma(\lambda^i)$, since the isothermal coordinates~$x^\alpha$ are introduced separately on each surface.

\begin{figure}[t]
\centering
\includegraphics[width=0.4\textwidth,page=4]{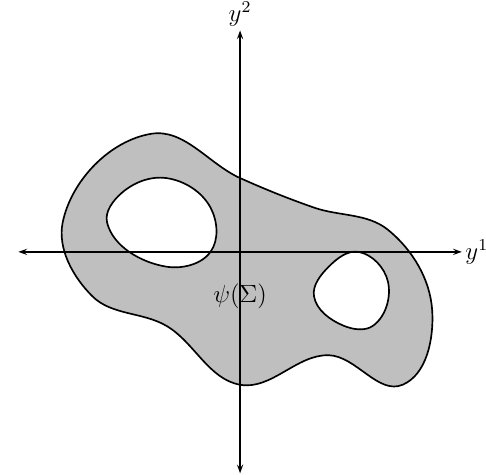}
\caption{When~$\Sigma$ does not have disk topology, we still assume it is planar, so that it can be embedded as a subregion~$\psi(\Sigma) \subset \mathbb{R}^2$ via a single coordinate chart~$\psi$.  This embedding will have ``holes'', and we extend the Jacobi operator~$J_\Sigma$ to the entire~$y$-plane including the interiors of these holes.  The interiors of these holes are part of the region~$\Omega_y$ on which the exterior problem~\eqref{eqs:exteriorholes} is posed.}
\label{fig:nondiskembedding}
\end{figure}

We assume that~$\Sigma$ is planar, and hence regardless of the metric~$\sigma_{\alpha\beta}$ on it we may introduce a coordinate system~$\{y^\alpha\}$ via a map~$\psi$ that embeds~$\Sigma$ as a subregion~$\psi(\Sigma) \subset \mathbb{R}^2$, as shown in Figure~\ref{fig:nondiskembedding} (because~$\Sigma$ need not have disk topology,~$\psi(\Sigma)$ need not be simply-connected).  If~$\Sigma$ has disk topology, we introduce the coordinate system~$\{x^\alpha\}$ exactly as above.  If~$\Sigma$ does not have disk topology, then it must have at least one ``hole''.  Let us now extend the Jacobi operator~$J_\Sigma$ to the entirety of the~$y$-plane, \textit{including the interiors of the holes}, by taking~$D_a$ to be flat and~$Q_{ij}$ to vanish outside of~$\psi(\Sigma)$, while~$\sigma_{\alpha\beta}$ is arbitrary (but known) outside of~$\psi(\Sigma)$, and~$\sigma_{\alpha\beta} = \delta_{\alpha\beta}$ outside of a region containing~$\psi(\Sigma)$.  The reason for taking~$\sigma_{\alpha\beta} = \delta_{\alpha\beta}$ outside of a region containing~$\psi(\Sigma)$, rather than \textit{everywhere} outside~$\psi(\Sigma)$, is to ensure that~$\sigma_{\alpha\beta}$ is differentiable at~$\partial\psi(\Sigma)$ (i.e.~we allow for a differentiable transition from~$\sigma_{\alpha\beta}|_{\partial \psi(\Sigma)}$ to~$\delta_{\alpha\beta}$).

We again consider the ``exterior'' problem~\eqref{eqs:exterior}:
\begin{subequations}
\label{eqs:exteriorholes}
\begin{align}
J_\Sigma \eta_\xi^i = 0 \mbox{ on } \Omega_y \equiv \mathbb{R}^2 \setminus \psi(\Sigma), \\
e^{-(y^1 + i y^2)\xi} \eta_\xi^i - \bar{\eta}^i \to 0 \mbox{ at large } |y|, \\
\left(\eta_\xi^i|_{\partial \psi(\Sigma)}, N^b D_b \eta^i_\xi|_{\partial \psi(\Sigma)}\right) \in \Ccal_\Sigma,
\end{align}
\end{subequations}
where now the region~$\Omega_y$ is disconnected because it contains the interiors of the holes.  To study solutions of this exterior problem, it will in fact be easier to study the problem on the entire plane:
\begin{subequations}
\label{eqs:fullproblem}
\begin{align}
J_\Sigma \eta_\xi^i = 0 \mbox{ on } \mathbb{R}^2, \\
e^{-(y^1 + i y^2)\xi} \eta_\xi^i - \bar{\eta}^i \to 0 \mbox{ at large } |y|;
\end{align}
\end{subequations}
note that the topology of~$\Sigma$ has now disappeared.  It is clear that any solution to~\eqref{eqs:fullproblem} gives rise to a solution of~\eqref{eqs:exteriorholes} by simply restricting to~$\Omega_y$.  Moreover, if the solution to~\eqref{eqs:fullproblem} is unique (for a given~$\xi$ and~$\bar{\eta}^i$), then so is the solution to~\eqref{eqs:exteriorholes}.  To see this, proceed by contradiction and suppose that~\eqref{eqs:exteriorholes} has two distinct solutions~$(\eta_\xi^i)_1$ and~$(\eta_\xi^i)_2$; then each one can be extended to the entire plane by solving the interior problem~$J_\Sigma \eta_\xi^i = 0$ on~$\psi(\Sigma)$ subject to the boundary conditions~$\eta_\xi^i|_{\partial \psi(\Sigma)} = (\eta_\xi^i)_1|_{\partial \psi(\Sigma)}$ and~$\eta_\xi^i|_{\partial \psi(\Sigma)} = (\eta_\xi^i)_2|_{\partial \psi(\Sigma)}$, respectively (this interior problem has a unique solution by virtue of the fact that~$\Sigma$ is weakly stable).  Stitching these interior solutions to the exterior ones~$(\eta_\xi^i)_{1,2}$ thus gives two distinct solutions to~\eqref{eqs:fullproblem}, which contradicts the assumption that the solution to~\eqref{eqs:fullproblem} is unique.  Hence we conclude that the solution to~\eqref{eqs:exteriorholes} must be unique if the solution to~\eqref{eqs:fullproblem} is.

We now introduce the map~$\Phi$ from the~$y^\alpha$ coordinates to the isothermal coordinates~$x^\alpha$; this map is unique if we require~$x^\alpha(y) \to y^\alpha$ at large~$|y|$, which we do~\cite{Ahlfors}.  Denote~$z = x^1 + i x^2$,~$\bar{z} = x^1 - i x^2$; then from~\cite{Bao_2019}, the problem boils down to
\begin{subequations}
\label{eqs:exteriorisothermal}
\begin{align}
F_\xi \delta \eta^i_\xi = -F_\xi \bar{\eta}^i \mbox{ on } \mathbb{C}, \\
\delta \eta^i_\xi \to 0 \mbox{ at large } |z|,
\end{align}
\end{subequations}
where we have defined
\be
\delta \eta^i_\xi(z) \equiv e^{-\xi z} \eta^i_\xi(z) - \bar{\eta}^i
\ee
and
\be
F_\xi u^i \equiv e^{2\phi} e^{-\xi z} J_\Sigma \left(e^{\xi z} u^i \right).
\ee
$F_\xi$ is a second-order uniformly elliptic differential operator which is given simply by~$F_\xi = \partial_z \partial_{\bar{z}} + \xi \partial_{\bar{z}}$ in the exterior region~$\Phi(\Omega_y)$ where~$D_a$ is flat and~$Q_{ab}$ vanishes.

Again as in~\cite{Bao_2019}, we claim that the problem~\eqref{eqs:exteriorisothermal} has a unique solution (for given~$\xi$,~$\bar{\eta}^i$) because~$F_\xi$ is a uniformly elliptic operator; the solution can be obtained by integrating~$-F_\xi \bar{\eta}^i$ against an appropriate Dirichlet Green's function.  Moreover, the behavior of the solution for large~$|\xi|$ is simple: when~$|\xi|$ is large, the problem~\eqref{eqs:exteriorisothermal} becomes
\bea
\left(\partial_z \partial_{\bar{z}} + \xi \partial_{\bar{z}}\right) \delta \eta_\xi^i + \cdots = 0 \mbox{ on } \mathbb{C}, \\
\delta \eta^i_\xi \to 0 \mbox{ at large } |z|,
\eea
where the ellipses denote~$\Ocal(\xi^0)$ terms with no derivatives.  Hence we find that for large~$\xi$,~$\delta \eta_\xi^i$ (for each~$i$) must approach a holomorphic function:~$\partial_{\bar{z}} \delta \eta^i_\xi \sim 1/\xi \to 0$.  But since~$\delta \eta_\xi^i$ must be regular everywhere (since it is the solution of a uniformly elliptic differential equation with regular coefficients and sources),~$\delta \eta_\xi^i$ must always be bounded, and hence as~$|\xi| \to \infty$,~$\delta \eta_\xi^i$ must approach a bounded entire function.  By Liouville's theorem, the only bounded entire functions are constants, and hence since~$\delta \eta_\xi^i \to 0$ at large~$|z|$, we find that~$\delta \eta_\xi^i$ vanishes \textit{everywhere} in the complex plane as~$|\xi| \to \infty$\footnote{In~\cite{Bao_2019}, the analogous argument is made somewhat too quickly: there it is claimed that because~$\partial_{\bar{z}} \delta \eta_\xi^i \to 0$ as~$|\xi| \to \infty$, the individual derivatives~$\partial \delta \eta_\xi^i/\partial x^\alpha$ must vanish as well.  This does not hold locally, as implied in~\cite{Bao_2019}; it only follows globally by virtue of Liouville's theorem because~$\delta \eta_\xi^i$ is bounded.  The importance of Liouville's theorem was not noted in the argument of~\cite{Bao_2019}, but the final conclusions remain correct.}.

We have therefore found that~\eqref{eqs:exteriorisothermal} admits a unique solution on the complex plane which vanishes as~$|\xi| \to 0$.  This then implies that the exterior problem~\eqref{eqs:exteriorholes} admits a unique solution in the region~$\Omega_y$, and this solution has the property
\be
\label{eq:etaxilimit}
\eta^i_\xi \to \bar{\eta}^i e^{(x^1(y) + i x^2(y))\xi} \mbox{ as } |\xi| \to \infty.
\ee
This is sufficient to deduce the existence of a unique set of isothermal coordinates on~$\Sigma$ that preserve the boundary structure, regardless of the metric on~$\Sigma$.  To see this, note that the exterior boundary problem~\eqref{eqs:exteriorholes} is insensitive to the metric on~$\Sigma$ except through the Cauchy data~$\Ccal_\Sigma$; hence for a given~$\xi$ and~$\bar{\eta}^i$,~$\eta_\xi^i$ in the exterior region~$\Omega_y$ is fixed by boundary data.  But from~\eqref{eq:etaxilimit} it follows that the isothermal coordinates~$x^\alpha$ are related to the~$y^\alpha$ (which are arbitrary coordinates on~$\Sigma$ independent of the metric) by\footnote{Technically the approach being described here, which we thank Adrian Nachman for pointing out to us, may run into subtleties involving the complex logarithm; but the same conclusion can be reached via the more rigorous, albeit less intuitive, analysis used in~\cite{AleBal17,Bao_2019}.}
\be
x^1(y) + i x^2(y) = \lim_{\xi \to \infty} \frac{1}{\xi} \ln \left(\frac{\eta_\xi^i(y)}{\bar{\eta}^i}\right) \quad \forall y \in \Omega_y, \quad \forall i.
\ee
The objects on the right-hand side can be obtained entirely from boundary data at any point~$y \in \Omega_y$, including at the boundary~$\partial \psi(\Sigma)$.  Hence the transformation to isothermal coordinates~$\{x^1(y),x^2(y)\}$ of all points on the boundary of~$\Sigma$ is fixed entirely by boundary data.  As claimed, we may therefore introduce the same set of isothermal coordinates for any two metrics on~$\Sigma$ with the same boundary data, meaning that the coordinate system~$\{x^\alpha,\lambda^i\}$ is a good one in which to work even for two different bulk metrics, as long as they share the same boundary data.

\subsection{Local Reconstruction}
\label{subsec:local}

\begin{figure}[t]
\centering
\subfloat[][]{
\includegraphics[width=0.3\textwidth,page=5]{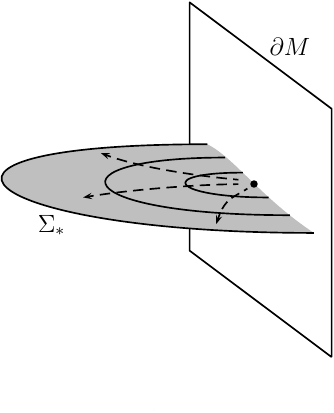}
\label{subfig:oldevolution}
}%
\hspace{1cm}
\subfloat[][]{
\includegraphics[width=0.28\textwidth,page=6]{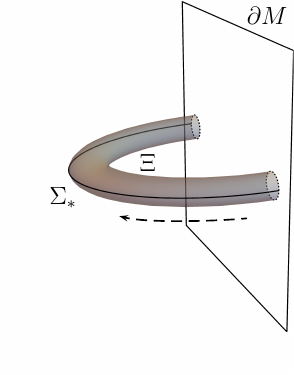}
\label{subfig:newevolution}
}
\caption{\protect\subref{subfig:oldevolution}: In the approach of~\cite{Bao_2019} described in Section~\ref{sec:review}, the conformal factor on some particular surface~$\Sigma_*$ of the foliation~$\Sigma(\lambda^i)$ is obtained by evolving~\eqref{eq:hyperbolic} inwards from the boundary along the~$\lambda^i$, indicated by dotted lines.  This evolution requires the foliation~$\Sigma(\lambda^i)$ to be continuous from~$\Sigma_*$ to the marked point on the boundary.  \protect\subref{subfig:newevolution}: Here we instead recover the conformal factor on~$\Sigma_*$ by treating~\eqref{eq:hyperbolic} as a boundary-value problem on a three-dimensional tube~$\Xi$ containing~$\Sigma_*$ constructed from a closed cycle of the~$\lambda^i$.  This tube can live in an arbitrarily small neighborhood of~$\Sigma_*$, eliminating the need for a foliation that transitions continuously from~$\Sigma_*$ to a point on the boundary.}
\label{figs:evolution}
\end{figure}

Next, let us discuss the removal of requirement~\ref{assump:point}: that is, the requirement that the foliation~$\Sigma(\lambda^i)$ shrink to a point on~$\partial M$ in an appropriate limit of the~$\lambda^i$.  From Section~\ref{sec:review}, recall that this assumption was needed in the argument of~\cite{Bao_2019} in order to fix the conformal factor via the extremality conditions
\be
\label{eq:hyperbolic}
\sum_{\alpha=1}^2 (\partial_{\alpha} f_{\alpha i} + 2f_{\alpha i}\partial_{\alpha}\phi)-2\partial_i\phi = 0 \qquad \forall i = 3, \ldots, d,
\ee
where the~$f_{\alpha i}$ depend only on~$g^{ij}$ and~$g^{\alpha i}$ and hence are uniquely fixed by the boundary data.  For each~$i$,~\eqref{eq:hyperbolic} can be interpreted as a hyperbolic equation for~$\phi$ with~$\lambda^i$ playing the role of ``time'', and hence we may evolve it inwards from the boundary as shown in Figure~\ref{subfig:oldevolution}.  In order to obtain the conformal factor on some particular surface~$\Sigma_* \equiv \Sigma(\lambda_*^i)$, we must therefore be able to evolve continuously from a point on the boundary to~$\Sigma_*$.

On the other hand, we could instead try to treat~\eqref{eq:hyperbolic} as a boundary-value problem.  To do so, we first need to construct some three-dimensional surface on which we can solve~\eqref{eq:hyperbolic} for~$\phi$.  A natural way to perform this construction is to consider a tubular neighborhood of~$\Sigma_*$, which by assumption will be foliated by the surfaces~$\Sigma(\lambda^i)$.  Within this tubular neighborhood, we identify some one-dimensional closed cycle in the~$\lambda^i$ parameters\footnote{Note that the existence of such a closed cycle in the~$\lambda^i$ is only guaranteed when~$d-2$ is at least two; that is, the~$\Sigma(\lambda^i)$ are at least codimension-two.} that starts and ends at~$\lambda_*^i$; the corresponding surfaces will form a three-dimensional tube containing~$\Sigma_*$, as shown in Figure~\ref{subfig:newevolution}.  We call this tube~$\Xi$.  Without loss of generality, we may redefine the~$\lambda^i$ so that, say, the cycle that defines this tube corresponds to varying~$\lambda^3$ while keeping all the other~$\lambda^i$ fixed.  Let us make this choice from now on; then the~$i = 3$ component of~\eqref{eq:hyperbolic} is a scalar first-order partial differential equation on~$\Xi$:
\be
\label{eq:hyperbolictube}
\sum_{\alpha=1}^2 (\partial_{\alpha} f_{\alpha 3} + 2f_{\alpha 3}\partial_{\alpha}\phi)-2\partial_3 \phi = 0.
\ee

Now, consider two different metrics~$g_1$ and~$g_2$ in the region~$\Rcal$ with the same boundary data.  By the arguments in the previous sections, in the coordinate system~$\{x^\alpha,\lambda^i\}$ the inverse metric components~$g^{ij}$ and~$g^{\alpha i}$ of these two metrics agree, and so they can only differ in the conformal factors~$\phi_1$ and~$\phi_2$.  Since both of these conformal factors must satisfy~\eqref{eq:hyperbolictube} (with the same~$f_{\alpha 3}$, since these depend only on~$g^{ij}$ and~$g^{\alpha i}$), the difference~$\delta \phi \equiv \phi_1 - \phi_2$ obeys
\be
\label{eq:deltaphi}
\left(\sum_{\alpha=1}^2 f_{\alpha 3}\partial_{\alpha} - \partial_3 \right) \delta \phi = 0.
\ee
Because the boundary data of~$g_1$ and~$g_2$ match, we must have~$\delta \phi|_{\partial M} = 0$; we now argue that the only solution of~\eqref{eq:deltaphi} with this boundary condition is just the trivial one~$\delta \phi = 0$.  This can be done easily using the method of characteristics:~\eqref{eq:deltaphi} is of the form
\be
\sum_{A = 1}^3 v^A \partial_A \delta \phi = 0
\ee
for~$v^A = (f_{13}, f_{23}, -1)$, where~$A = 1, 2, 3$ indexes the coordinate system~$\{x^1, x^2, \lambda^3\}$ on~$\Xi$.  Now, consider an integral curve~$\gamma^A(t)$ of the vector field~$v^A$ (that is, a curve with tangent~$d \gamma^A/dt = v^A$); such curves are the \textit{characteristics} of~\eqref{eq:deltaphi}.  Along characteristics,
\be
\frac{d \delta \phi(\gamma(t))}{dt} = 0,
\ee
and hence any solution of~\eqref{eq:deltaphi} is constant on characteristics.  Since~$\delta \phi|_{\partial M} = 0$, we immediately conclude that~$\delta \phi$ must vanish at all points in~$\Xi$ that can be reached from~$\partial \Xi$ along characteristics.  If all points on~$\Xi$ can be reached in such a way, then we are done.  If there are points on~$\Xi$ that \textit{cannot} be reached along characteristics from~$\partial \Xi$, then such points must lie on characteristics contained entirely within~$\Sigma$.  Typically such characteristics will start at repulsors and end at attractors, and the various basins of repulsion and attraction on~$\Xi$ will be connected.  But since~$\delta \phi|_{\partial M} = 0$,~$\delta \phi$ must vanish at any repulsors or attractors reached along characteristics from~$\partial \Xi$.  Continuity of~$\delta \phi$ then requires that~$\delta \phi$ vanish along all characteristics starting and ending at such repulsors and attractors, and since all the basins are connected, we conclude that~$\delta \phi$ must in fact vanish everywhere.

In very fine-tuned special cases, there may be families of characteristics that do not start or end at repulsors or attractors (e.g.~if the characteristics form closed cycles).  In such a case, the argument above does not work.  But the fine-tuning required for engineering such a scenario is highly non-generic; an arbitrarily small perturbation to~$v^A$ can restore the decomposition of~$\Xi$ into a set of connected basins of repulsion and attraction, and presumably such a deformation can be engineered by an appropriate deformation of~$\Xi$.  Hence we will not concern ourselves futher with such special cases.  With this understanding in mind, we therefore conclude that the zero solution is the unique solution to~\eqref{eq:deltaphi}, and hence that the two conformal factors~$\phi_1$ and~$\phi_2$ must in fact agree.  Hence the metrics~$g_1$ and~$g_2$ with the same boundary data are in fact the same.

\section{Genericity of the Foliation Condition}
\label{sec:foliation}

We have now substantially weakened Assumption~\ref{assump:disk} and removed Assumption~\ref{assump:point}, no longer requiring that the surfaces~$\Sigma(\lambda^i)$ have disk topology nor that they degenerate to a point on the boundary.  Of the remaining assumptions, the stability requirement~\ref{assump:stable} is extremely generic, so the most contentful remaining assumption is~\ref{assump:foliation}: that the~$\Sigma(\lambda^i)$ foliate some portion~$\Rcal$ of~$M$.  The purpose of this section is to explore the genericity of this assumption in~$d = 4$ AlAdS spacetimes.

\subsection{Stationary Spacetimes}
\label{subsec:stationary}

As discussed in~\cite{Bao_2019}, the foliation condition holds identically in stationary spacetimes obeying the null curvature condition (NCC)\footnote{The null curvature condition states that~$R_{ab} k^a k^b \geq 0$ everywhere in~$M$, where~$k^a$ is any null vector and~$R_{ab}$ is the Ricci tensor of~$g_{ab}$.}, as can be seen as follows.  Consider first an arbitrary boundary-anchored two-dimensional spacelike extremal surface~$\Sigma_*$ in such a spacetime anchored to an achronal region~$B$ on~$\partial M$ (that is,~$\partial B = \partial \Sigma_*$).  Now suppose that the region~$B$ is deformed into a one-parameter family~$B(s)$ of monotonically shrinking regions: that is,~$B(0) = B$ and~$B(s_2) \subset B(s_1)$ for any~$s_2 > s_1$.  There will be a corresponding one-parameter family~$\Sigma(s)$ of bulk extremal surfaces anchored to~$\partial B(s)$ with~$\Sigma(0) = \Sigma_*$.  As shown in~\cite{EngFis19} (and earlier in~\cite{EngWal13,Wal12} under additional assumptions), the null curvature condition ensures that~$\Sigma(s_2)$ always lies to the outside of~$\Sigma(s_1)$ for all~$s_2 > s_1$; in other words, the surfaces~$\Sigma(s)$ move monotonically ``outwards'' as the boundary region~$B$ is shrunk\footnote{When the surfaces~$\Sigma(s)$ are HRT surfaces in a holographic spacetime, this feature is known as entanglement wedge nesting and follows from causality of the dual CFT, but holography does not need to be invoked to deduce this nesting as long as the NCC is assumed in the bulk.}.  Consequently, the family~$\Sigma(s)$ sweeps out a three-dimensional achronal hypersurface~$N$ which is foliated by the~$\Sigma(s)$. 

Now, if the bulk spacetime is stationary, there exists a time-translation Killing vector field~$\xi^a$.  Let us evolve the one-parameter family of boundary regions~$B(s)$ into a two-parameter family~$B(s,t)$ by time-translating along this Killing field: i.e.,~$B(s,t)$ is obtained from~$B(s)$ by evolving a Killing parameter~$t$ along the integral curves of~$\xi^a$.  Since~$\xi^a$ is a Killing field, it follows that there exists a corresponding two-parameter family of extremal surfaces~$\Sigma(s,t)$ anchored to~$\partial B(s,t)$ such that~$\Sigma(s,t)$ is also obtained by time-evolving~$\Sigma(s)$ along the integral curves of~$\xi^a$.  For each fixed~$t$, the surface~$N(t)$ swept out by the~$\Sigma(s,t)$ as~$s$ is varied is again a three-dimensional achronal hypersurface foliated by the~$\Sigma(s,t)$ (with~$N(0) = N$).  Moreover, since~$\xi^a$ is timelike, it is not tangent to any of the~$N(t)$, and hence the various hypersurfaces~$N(t)$ foliate some neighborhood of the original surface~$\Sigma_*$.  Since each~$N(t)$ is foliated by the~$\Sigma(s,t)$, we therefore conclude that this construction ensures the existence of a foliation of extremal surfaces~$\Sigma(s,t)$ in a neighborhood of~$\Sigma_*$.

In a holographic context, one might be concerned that perhaps the~$\Sigma(s,t)$ will cease to be HRT surfaces as the parameters~$s$ and~$t$ are varied (if, for instance, there are other extremal surfaces anchored to the same boundary region~$B(s,t)$ that end up having smaller area than the~$\Sigma(s,t)$).  But since the construction is local, if~$\Sigma_*$ is an HRT surface, we can always find a sufficiently small neighborhood of~$\Sigma_*$ foliated by HRT surfaces unless~$\Sigma_*$ itself lies right on a cusp of a phase transition: that is, if there is another extremal surface different from~$\Sigma_*$ also anchored to~$\partial \Sigma_*$ and with the same area as~$\Sigma_*$.  Such a choice of~$\Sigma_*$ is highly non-generic, and hence we conclude that for generic HRT surfaces in a stationary spacetime, the foliation condition is always obeyed.

\subsection{Case Study in a Non-Stationary Spacetime: AdS-Vaidya}
\label{subsec:Vaidya}

Since the foliation condition is always satisfied in stationary spacetimes, one might expect that it should still be satisfied for sufficiently small perturbations theoreof.  However, sufficiently dynamical spacetimes may violate it (specifically, while entanglement wedge nesting holds in any spacetime satisfying the NCC, and hence the achronal surface~$N$ described in the previous subsection is always foliated by the~$\Sigma(s)$, time-evolution of the boundary regions~$R(s)$ need not correspond to a uniform time-evolution of the corresponding surfaces~$N(t)$ once stationarity is lost).  To explore this possibility, let us therefore investigate a tractable example of a dynamical spacetime: planar AdS-Vaidya.  We will find that even in the highly-dynamical region of the geometry, the foliation condition is obeyed, suggesting that even highly dynamical spacetimes will not generically cause it to be violated.

The AdS-Vaidya metric we consider is the spacetime sourced by a the collapse of a plane of null matter in AdS.  In ingoing Eddington-Finkelstein coordinates, it is given by
\be
\label{eq:AdSVaidya}
ds^2=\frac{L^2}{z^2}(-f(z,v)dv^2-2dvdz + d\rho^2 + \rho^2 \, d\theta^2),
\ee
where~$f(v,z)=1-m(v)z^3$ with~$m(v)$ the profile of the infalling matter,~$L$ is the AdS scale, and we have written the flat spatial boundary metric in polar coordinates~$(\rho,\theta)$.  Starting in Poincar\'e AdS corresponds to~$m(v = -\infty) = 0$, and the final horizon size~$z_h$ is set by the late-time value~$m(v = +\infty) = 1/z_h^3$.   In what follows, we will set~$z_h=1$ and we will take the matter profile to be given by
\be
m(v) = \begin{cases}
        0, & v\leq -\pi T/2, \\
        \frac{1}{2}\left(1+\sin(v/T)\right), & -\pi T/2 \leq v\leq \pi T/ 2,\\
        1, & \text{otherwise},
        \end{cases}
\ee
where~$T$ is a tunable parameter that sets the thickness of the matter shell; this profile corresponds to starting with Poincar\'e AdS and forming a planar black hole by injecting some null energy flux from the boundary during a time window $v \in (-\pi T/2, \pi T/2)$.

The dynamical portion of the spacetime corresponds to the mass shell in the region~$v \in (-\pi T/2, \pi T/2)$, and hence we are interested in whether there are regions of this shell that can be foliated by (portions of) HRT surfaces (the pure AdS and Schwarzschild regions are stationary, so foliations of HRT surfaces can always be found in those regions).  To proceed, we will consider spherically-symmetric HRT surfaces anchored to boundary circles of constant~$v$ and~$\rho$.  It will then be natural to take~$\rho$ and~$\theta$ as coordinates on our HRT surfaces, which we will parametrize as~$v = V(\rho)$,~$z = Z(\rho)$.  An analysis of such surfaces was performed in~\cite{Liu:2013iza,Liu:2013qca}, which we now briefly review.

The area functional of these spherically-symmetric surfaces is given by
\be
A = K \int_0^R d\rho \, \frac{\rho}{z^2} \sqrt{Q}, \mbox{ where }  Q \equiv 1-2V'Z'-f(Z,V)V'^2;
\ee
primes denote derivative with respect to~$\rho$;~$R$ sets the size of the boundary circle~$\rho = R$ to which the surface is anchored; and $K$ is a constant not important for our purposes.  The equations of motion for~$V$ and~$Z$ are then obtained by extremizing with respect to them:
\begin{subequations}
\label{eqs:EOM}
\begin{align}
    \frac{Z^2\sqrt{Q}}{\rho} \left(\frac{\rho V'}{Z^2\sqrt{Q}}\right)' &= \frac{2Q}{Z}+\frac{1}{2}\frac{\partial f}{\partial z} \, V'^2, \\
    \frac{Z^2\sqrt{Q}}{\rho} \left(\frac{\rho(Z'+fV')}{Z^2\sqrt{Q}}\right)' &= \frac{1}{2}\frac{\partial f}{\partial v} \, V'^2.
\end{align}
\end{subequations}
These equations can be solved numerically for the HRT surfaces.  The natural boundary conditions are
\be
Z(R)= 0,\quad V(R)= v_0,\quad Z'(0)=0,\quad V'(0)=0,
\ee
where we take the HRT surfaces to be anchored on the boundary at time~$v = v_0$.  For obtaining numerical solutions of~\eqref{eqs:EOM}, it is in fact simpler to exchange the disk radius~$R$ and boundary time~$v_0$ with the location~$(v_\mathrm{turn}, z_\mathrm{turn})$ of the turning point.  In other words, we impose the ``initial conditions''
\be
\label{eq:initialconds}
Z(0)= z_\mathrm{turn},\quad V(0)= v_\mathrm{turn},\quad Z'(0) = 0,\quad V'(0) = 0;
\ee
for a given choice of~$(v_\mathrm{turn}, z_\mathrm{turn})$, one can then read off the corresponding boundary values~$(R,v_0)$ from the solution to~\eqref{eqs:EOM}\footnote{In practice, since~\eqref{eqs:EOM} are singular at~$\rho = 0$, we actually impose initial conditions at a cutoff~$\rho = \eps$ for some sufficiently small~$\eps$.  With this modification,~$Z'(\eps)$ and~$V'(\eps)$ are chosen to be small but nonzero to be consistent with~\eqref{eq:initialconds}.}.

In the pure AdS region~$v < -\pi T/2$, the usual global Killing time~$t$ is given by~$t = v + z$, and hence in this region an HRT surface with turning point~$(v_\mathrm{turn}, z_\mathrm{turn})$ will lie on the constant-time slice~$t = t_\mathrm{turn} \equiv v_\mathrm{turn} + z_\mathrm{turn}$.  Because we are interested in HRT surfaces sensitive to the region of dynamical geometry, we take~$-z_\mathrm{turn} - \pi T/2 < v_\mathrm{turn} < -\pi T/2$ to guarantee that the HRT surface will cross the matter shell.  By varying~$z_\mathrm{turn}$ and~$v_\mathrm{turn}$, we obtain a two-parameter family of HRT surfaces, and can then investigate whether or not these surfaces provide a foliation of some region of the bulk.

\begin{figure}
\centering
\includegraphics[width=0.4\textwidth,page=7]{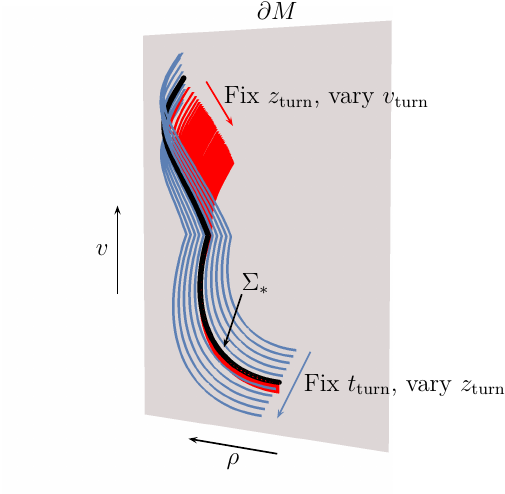}
\caption{HRT surfaces in planar AdS-Vaidya~\eqref{eq:AdSVaidya} with~$T = 1/150$; without loss of generality we have set the AdS length~$L = 1$.  Here the angular direction is suppressed, so each curve is really a topological disk that ``caps off'' at~$\rho = 0$.  The black surface~$\Sigma_*$ corresponds to taking~$z_\mathrm{turn} = 3.5$ and~$v_\mathrm{turn} = -2.21$; the red surfaces are then obtained by varying~$v_\mathrm{turn}$ while keeping~$z_\mathrm{turn}$ fixed, while the blue surfaces correspond to varying~$v_\mathrm{turn}$ and~$z_\mathrm{turn}$ in a way that keeps~$t_\mathrm{turn} = v_\mathrm{turn} + z_\mathrm{turn}$ fixed.  The deformations corresponding to the red and blue families of surfaces are linearly independent on the black surface, implying that a neighborhood of the black surface is foliated by boundary-anchored HRT surfaces.}
\label{fig:Vaidyafoliation}
\end{figure}

A typical result is shown in Figure~\ref{fig:Vaidyafoliation}, highlighting a particular choice of reference surface~$\Sigma_*$ and its deformation as~$z_\mathrm{turn}$ and~$v_\mathrm{turn}$ are varied.  The crucial feature to note is that varying~$z_\mathrm{turn}$ and~$v_\mathrm{turn}$ independently deforms~$\Sigma_*$ in two linearly independent directions (more precisely, the deviation vectors~$(\partial_{v_\mathrm{turn}})^a$ and~$(\partial_{t_\mathrm{turn}})^a$ on~$\Sigma_*$ are linearly independent, which can be seen from Figure~\ref{fig:Vaidyafoliation} by noting that on~$\Sigma_*$, the tangent vector to the hypersurface swept out by the blue HRT surfaces is linearly independent from the tangent vector to the hypersurface swept out by the red HRT surfaces).  This behavior is common to all choices of~$(v_\mathrm{turn}, z_\mathrm{turn})$ we have studied; we were unable to find any instance in which a neighborhood of a (spherically-symmetric) HRT surface failed to be foliated by HRT surfaces.  Moreover, the particular choice of~$\Sigma_*$ exhibited in Figure~\ref{fig:Vaidyafoliation} in fact penetrates through the event horizon: in the Schwarzschild region~$v > \pi T/2$, the event horizon lies at~$z = 1$, and it is straightforward to check that for the black surface shown in Figure~\ref{fig:Vaidyafoliation},~$Z(\rho) > 1$ shortly after crossing the matter shell into the Schwarzschild region -- specifically, the matter shell is crossed at~$Z(\rho) \approx 1.3$.  Hence, in this particular example, not only do we fail to find violations of the foliation condition even for surfaces that enter the highly-dynamical matter shell, but we also find no violation of the foliation condition for HRT surfaces that enter the event horizon.

\subsection{An Infinitesimal Foliation Condition}
\label{subsec:inffoliation}

While it is encouraging that the foliation condition appears to always be satisfied in the planar AdS-Vaidya geometry (at least for the spherically symmetric surfaces we considered), the analysis above suggests that perhaps more can be said if we restrict ourselves to only looking for a foliation in an \textit{infinitesimal} neighborhood of some reference surface~$\Sigma_*$. That is, what if we require that there exist a two-parameter family of \textit{infinitesimal} deformations of~$\Sigma_*$ that ``foliate'' its neighborhood?  Since infinitesimal deformations of~$\Sigma_*$ obey the Jacobi equation, the sense in which the existence of an infinitesimal foliation is to be understood is to require that on~$\Sigma^*$, there exist two linearly independent vector fields~$\eta_1^a$ and~$\eta_2^a$ satisfying the Jacobi equation~\eqref{eq:Jacobi}.  In other words, for a four-dimensional spacetime, consider modifying condition~\ref{assump:foliation} to
\begin{enumerate}[label={\arabic*$'$}]
	\item Given a boundary-anchored, spacelike, two-dimensional extremal surface~$\Sigma_*$, there exist two linearly independent vector fields~$\eta_1^a$,~$\eta_2^a$ such that~$J_{\Sigma_*} \eta_{1,2}^a = 0$, where~$J_{\Sigma_*}$ is the Jacobi operator of~$\Sigma_*$. \label{assump:inffoliation}
\end{enumerate}
This condition is automatically satisfied if the finite foliation condition~\ref{assump:foliation} is (since if the family~$\Sigma(\lambda^i)$ specified by condition~\ref{assump:foliation} exists, then by definition the deviation vectors~$(\partial_{\lambda^1})^a$ and~$(\partial_{\lambda^2})^a$ must satisfy condition~\ref{assump:inffoliation} on all the~$\Sigma(\lambda^i)$), but the converse need not be true.  Hence the infinitesimal foliation condition~\ref{assump:inffoliation} can be thought of as a necessary condition for our argument to be applicable, but not a sufficient one.  Nevertheless, because~\ref{assump:inffoliation} is now a statement about properties of the Jacobi operator~$J_{\Sigma_*}$, we might hope there is more that can be said.

Unfortunately we will not be able to make any concrete statements; the purpose of this subsection will instead be to formulate the condition more precisely and to draw some exploratory connections between the infinitesimal foliation condition, cooperative elliptic systems, and positivity of elliptic operators.

To that end, we note that for a spacelike codimension-two surface~$\Sigma_*$ in a Lorentzian spacetime, we may introduce two independent null vectors~$k^a$ and~$\ell^a$ normal to~$\Sigma_*$, where we take~$k^a$ to be future-directed and~$\ell^a$ to be past-directed.  Decomposing an arbitrary vector normal to~$\Sigma_*$ as~$\eta^a = \alpha k^a + \beta \ell^a$, the Jacobi equation~$J \eta^a = 0$ becomes the system of elliptic equations
\be
\label{eq:Jacobidecomp}
\left[ \begin{pmatrix} J_+ & 0 \\ 0 & J_- \end{pmatrix} - \begin{pmatrix} 0 & Q_{kk} \\ Q_{\ell\ell} & 0 \end{pmatrix}
\right] \begin{pmatrix} \alpha \\ \beta \end{pmatrix} = \begin{pmatrix} 0 \\ 0 \end{pmatrix},
\ee
where
\bea
J_{\pm} &\equiv -D^2\mp 2\chi^a D_a-(|\chi|^2\pm D_a\chi^a+Q_{k\ell}), \\
\chi_a &\equiv \ell^b D_a k_b,
\eea
where as above~$D_a$ is the covariant derivative on~$\Sigma_*$ and~$Q_{ab}$ is defined in~\eqref{subeq:JacobiQ}; see e.g.~\cite{EngFis19} for details on this decomposition.  Now, the NCC implies that~$Q_{kk}$ and~$Q_{\ell\ell}$ are both non-negative; in such a case, elliptic systems of the form~\eqref{eq:Jacobidecomp} are known as \textit{cooperative elliptic systems}.  They obey several useful properties; for instance, if~$\alpha$ and~$\beta$ are both non-negative at~$\partial \Sigma_*$, then they must be non-negative everywhere on~$\Sigma_*$ as well~\cite{Swe92}.  Since non-negative~$\alpha$ and~$\beta$ correspond to an outwards-pointing~$\eta^a$, this is the statement alluded to in Section~\ref{subsec:stationary} above that entanglement wedge nesting follows from the NCC.

Now, the infinitesimal foliation condition~\ref{assump:inffoliation} requires that there exist two linearly independent deviation vector fields~$\eta_{1,2}^a$ on~$\Sigma_*$.  Assuming the NCC, entanglement wedge nesting can be invoked to immediately deduce the existence of an everywhere-spacelike deviation vector~$\eta_s^a$: just solve~\eqref{eq:Jacobidecomp} with~$\alpha$ and~$\beta$ strictly positive on~$\partial \Sigma_*$, yielding strictly positive~$\alpha$ and~$\beta$ on~$\Sigma_*$.  If we could deduce the existence of another everywhere-\textit{timelike} deviation vector, then we would be done.

We do not know of a way to deduce the existence of such a vector (or of conditions necessary for its existence), and it is for this reason that we cannot give a precise alternative formulation of the infinitesimal foliation condition.  But we can rephrase the question in a potentially more illuminating way as follows.  A timelike deviation vector would correspond to an everywhere-negative~$\beta$ and an everywhere-positive~$\alpha$ (or vice versa), so let us define~$\tilde{\beta} = -\beta$; then~\eqref{eq:Jacobidecomp} becomes
\be
\label{eq:Jacobidecomptimelike}
\left[ \begin{pmatrix} J_+ & 0 \\ 0 & J_- \end{pmatrix} + \begin{pmatrix} 0 & Q_{kk} \\ Q_{\ell\ell} & 0 \end{pmatrix}
\right] \begin{pmatrix} \alpha \\ \tilde{\beta} \end{pmatrix} = \begin{pmatrix} 0 \\ 0 \end{pmatrix}.
\ee
This new system is not cooperative, so it need not \textit{preserve the positive cone}: that is, solutions with positive~$\alpha$ and~$\tilde{\beta}$ on~$\partial \Sigma_*$ need not be positive everywhere in~$\Sigma_*$.  The question we are asking is what conditions are necessary in order to preserve \textit{part} of the positive cone, so that there exists at least \textit{some} solution to~\eqref{eq:Jacobidecomptimelike} with everywhere-positive~$\alpha$ and~$\tilde{\beta}$.  Such a question has been studied in e.g.~\cite{MitSwe95} and references therein, but unfortunately to our knowledge there are no general results that can be straightforwardly mapped to intuitive properties of the Jacobi operator.  We do note, however, that if~$Q_{kk}$ and~$Q_{\ell\ell}$ both vanish, then~\eqref{eq:Jacobidecomptimelike} becomes identical to~\eqref{eq:Jacobidecomp}, and hence it too must have a solution with everywhere-positive~$\alpha$ and~$\tilde{\beta}$; thus a timelike deviation vector would exist when~$Q_{kk} = 0 = Q_{\ell\ell}$.  This observation makes it natural to work perturbatively in~$Q_{kk}$ and~$Q_{kk}$, and in~\cite{MitSwe95} (specifically Theorem~6.4; see also references therein) appropriate conditions on~$Q_{kk}$,~$Q_{ll}$, and~$J_\pm$ are given that ensure the existence of such solutions.  These conditions are rather technical and unilluminating so we do not give them here; we hope, however, that the connection to the question of preservation of the positive cone in noncooperative elliptic systems may be useful in future examinations of the foliation condition.

\section{Discussion}
\label{sec:disc}

In this Note, we have generalized the bulk reconstruction argument of~\cite{Bao_2019} in two ways.  First, we have softened the requirement that the extremal surfaces~$\Sigma(\lambda^i)$ have disk topology into merely the requirement that they be planar; second, we have removed the requirement that the family~$\Sigma(\lambda^i)$ must shrink to a point on the boundary.  These extensions apply in arbitrary bulk dimension and signature (though the~$\Sigma(\lambda^i)$ must still be spacelike and two-dimensional), but the context of primary interest is that of a four-dimensional AlAdS geometry, in which case the~$\Sigma(\lambda^i)$ can be interpreted as HRT surfaces.  In this context, our generalizations allow us to deduce uniqueness of the bulk metric in a local neighborhood of some particular HRT surface without requiring that the surface be continuously deformable to the boundary; in particular, this means that boundary entanglement entropy fixes the bulk metric in a neighborhood of an HRT surface even past a ``jump'' corresponding to a phase transition in the entanglement entropy.  Our generalizations also allow us to apply our argument to e.g.~HRT surfaces that connect two disconnected AdS boundaries through a two-sided black hole, deducing that the boundary entanglement entropy fixes the bulk metric in a portion of the black hole interior.

The strongest remaining assumption in the argument is that the~$\Sigma(\lambda^i)$ foliate some portion of the bulk.  While we have not weakened this assumption, we have explored its validity in more depth.  It is always satisfied in stationary spacetimes satisfying the null curvature condition, and we confirmed it is also satisfied in neighborhoods of spherically-symmetric HRT surfaces sensitive to the dynamical geometry region of AdS-Vaidya (including surfaces that enter the event horizon at sufficiently early times).  We have discussed a linearization of the foliation condition and its connection to the interesting structure of elliptic systems, in particular the question of preservation of the positive cone in noncooperative elliptic systems.

Many open questions remain, most of them already listed in~\cite{Bao_2019}.  These include, for instance, the question of how the argument can be made constructive in order to explicitly recover the bulk metric from boundary entropies; how the argument is affected under the inclusion of quantum corrections to the HRT formula~\cite{FauLew13,EngWal14}; and how to extend the argument to HRT surfaces in dimensions higher than four.  One might also investigate the applicability of our argument in various particular interesting choices of spacetime, such as the traversable wormhole constructions of \cite{Gao_2017,Maldacena_2017}.  How ``deep'' into the bulk can the metric be recovered from entanglement entropy, as per our argument?

A new additional question that presents itself is whether there is a way to make a more precise connection between the infinitesimal foliation condition discussion in Section~\ref{subsec:inffoliation} and properties of the Jacobi operator.  For example, is there a relationship between a given HRT surface~$\Sigma_*$ satisfying the infinitesimal foliation condition and its stability, as defined by the spectrum of its Jacobi operator~\cite{EngFis19}?  We leave these and other questions to future work.

\section*{Acknowledgements}

We thank Spyros Alexakis, Tracey Balehowsky, Ryan Hamerly, Cindy Keeler, Adrian Nachman, Philippe Sabella-Garnier, and Guido Sweers for useful discussions and comments.  N.B. is supported by the National Science Foundation under grant number 82248-13067-44-PHPXH, by the Department of Energy under grant number DE-SC0019380, and by the Computational Science Initiative at Brookhaven National Laboratory. C.C. is supported by the U.S. Department of Defense and NIST through the Hartree Postdoctoral Fellowship at QuICS, by the Simons Foundation as part of the It From Qubit Collaboration, and by the DOE Office of Science, Office of High Energy Physics, through the grant DE-SC0019380. SF acknowledges the support of the Natural Sciences and Engineering Research Council of Canada (NSERC), funding reference number SAPIN/00032-2015, and of a grant from the Simons Foundation (385602, AM). J.P. is supported in part by the Simons Foundation and in part by the Natural Sciences and Engineering Research Council of Canada.

\bibliographystyle{jhep}
\bibliography{all}

\end{document}